\documentclass{elsart}
\usepackage{graphicx}
\begin{document}
\runauthor{Madden}
\begin{frontmatter}
\title{Effects of massive star formation on the ISM of dwarf galaxies\thanksref{X}}
\author{Suzanne C. Madden}
\address{CEA, Saclay, Service d'Astrophysique, France, smadden@cea.fr}
\thanks[X]{\emph{Proceedings for The Interplay between Massive Stars and the ISM \\ New Astronomy Reviews, Eds. D. Schaerer \& R. Delgado-Gonzalez}}
\begin{abstract}
We are studying star formation effects on the properties of the ISM in
low metallicity environments using mid-infrared (MIR) and far-infrared
(FIR) observations of starbursting dwarf galaxies taken with the
Infrared Space Observatory (ISO) and the Kuiper Airborne Observatory
(KAO). Effects of the hard pervasive radiation field on the gas and
dust due to the dust-poor environments are apparent in both the dust
and gas components. From a 158 $\mu$m [CII] survey we find enhanced
I[CII]/FIR ratios in dwarf galaxies and I[CII]/I(CO) ratios up to 10
times higher than those for normal metallicity starburst galaxies.  We
consider MIR observations in understanding the star formation
properties of dwarf galaxies and constraints on the stellar
SED. Notably, the strong MIR [NeIII]/[NeII] ratios reveal the presence
of current massive stellar populations $<$ 5 My old in NGC~1569,
NGC~1140 and IIZw40. The MIR unidentified infrared bands (UIBs) are
weak, if present at all, as a general characteristic in low
metallicity environments, revealing the destruction of the smallest
carbon particles (e.g. PAHs) over large spatial scales. This is confirmed with our
dust modeling: mass fractions of PAHs are almost negligible compared
to the larger silicate grains emitting in the FIR as well as the small
carbon grains emitting in the MIR, which appear to be the source of
the photoelectric gas heating in these galaxies, in view of the [CII]
cooling.
\end{abstract}
\begin{keyword}
dwarf galaxies; dust;
photodissociation regions; ISO
\end{keyword}
\end{frontmatter}
\section{Introduction}
\typeout{SET RUN AUTHOR to \@runauthor}
To construct a comprehensive picture of a galaxy's history,
understanding the distribution of its energy budget is a fundamental
step.  For this we must consider observations covering several
characteristic wavelength regimes, thus, sampling the various
components of the interstellar medium (ISM). While the UV to NIR
wavelength continua give us relatively direct probes of the stellar
populations, this radiation is subject to varying amounts of
absorption before we view it. Some of this energy is absorbed by the
gas directly in HII regions or transferred to the gas in
photodissociation regions (PDRs) and reemitted as molecules, bands and
atomic ionic and recombination lines, from wavelengths covering the UV
to FIR and beyond. Some of the stellar energy is absorbed by the dust,
revealed through extinction, and reradiated in MIR to submillimeter
wavelengths as thermal emission.  Therefore, models of the ISM in
galaxies must consider these interdependent processes and be
self-consistent. Our knowledge of the wavelength window from the MIR
to the FIR has been limited by the low spatial and spectral resolution
provided by IRAS, and has been rather sketchy when it comes to
detailed studies of the ISM of individual galaxies. The Infrared Space
Observatory (ISO) \cite{kessler96} has been a recent turning point in
this effort, providing high spectral and spatial resolution and
unprecedented sensitivity in the MIR to FIR.  We are incorporating our
MIR and FIR observations in a study of the energy redistribution in
starburst galaxies to understand the effects of the star formation on
the surrounding gas and dust. Here we report the progress to date in
our study of star forming low-metallicity dwarf galaxies, which, in
the absence of major dynamical complications, allow us to `simplify'
model assumptions and the interpretation of observations.
\section{Far-infrared observations: the [CII] cooling line}
As an indirect probe of the star formation activity, we have obtained
KAO (Kuiper Airborne Observatory) and ISO observations of the 158
$\mu$m $^2P_{3/2} - ^2P_{1/2}$ far infrared [CII] fine structure line
emission in a sample of 15 dwarf galaxies \cite{jones97} with
metallicities ranging from 0.1 to 0.5 solar.  As the ionization
potential of carbon is 11.3 eV, less than that of HI, photons escaping
the HII regions, dissociate CO, and ionize carbon in the
photodissociation regions (PDRs) on the surfaces of nearby molecular
clouds exposed to the stellar UV radiation. The observed [CII]
intensity can be traced back to the radiation source due to the fact
that the UV photons heat the dust which emits thermal radiation in the
MIR to submillimeter wavelengths. Energetic electrons, ejected from
the dust through the photoelectric effect, heat the gas. The gas
subsequently cools via emission from molecules and atomic fine
structure lines, predominantly the 158 $\mu$m [CII] and the 63 $\mu$m
[OI] transitions in PDRs. There has been a long history of development
of PDR models which provide tools to differentiate physical
properties, such as density (n), radiation field strength (G$_{0}$)
and filling-factors in galaxies (see review and references in
\cite{hollenbach97}).
\subsection{[CII] Survey of Dwarf Galaxies}
\begin{figure*}
\includegraphics[width=13cm]{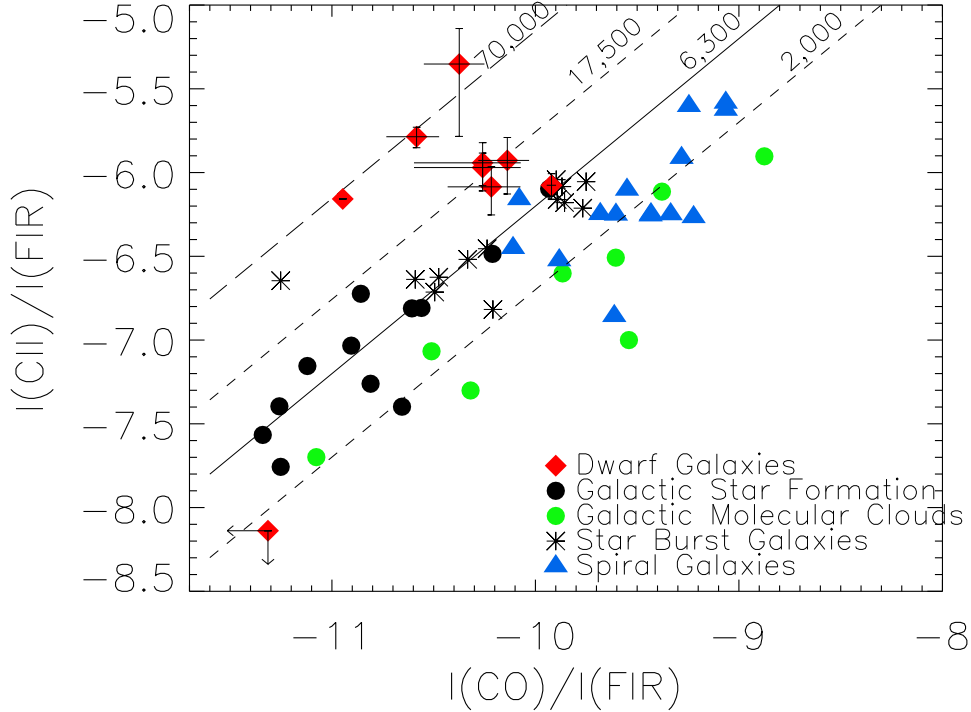}
\caption{[CII] survey results: comparing normal metallicity regions with low-metallicity galaxies. Lines of constant I[CII]/I(CO) ratios run diagonally across the plot and range from $\sim$ 2000 for quiescent galaxies and Galactic molecular cloud regions \cite{stacey91} up to $\sim$ 70,000 for some dwarf galaxies \cite{jones97}. The ratios of both axes are normalized to the local interstellar radiation field (1.3x10$^{-4}$ erg s$^{-1}$ cm$^{-2}$ sr$^{-1}$)}
\label{cii}
\end{figure*}

The ratio of I[CII]/I(CO) is a useful measure of the PDR emission
relative to the molecular core emission and is an indicator of the
degree of star formation activity in galaxies. Active galaxies have a
ratio of I[CII]/I(CO) $\sim$ 6300, which is 3 times greater than that
observed in more quiescent galaxies \cite{stacey91}.  Our [CII] survey
shows that for dwarf galaxies, this ratio ranges from 6000 to 70,000,
which is up to 10 times greater than those for normal metallicity
starburst galaxies (Figure \ref{cii}) \cite{jones97}.  We also observe
an overall enhancement in the I[CII]/FIR ratios (where FIR is defined
as the sum of the IRAS 60 and 100 $\mu$m bands) in these regions
compared to those in normal metallicity galaxies, which was also noted
in the LMC \cite{mochizuki94} \cite{poglitsch95}
\cite{pak98}.  The ratio of I[CII]/FIR is a direct measure of the
fraction of UV energy reemerging in the [CII] cooling line, and is
usually between 0.1\% and 1\% for normal metallicity galaxies
\cite{stacey91}, while we find up to 2\% for dwarf galaxies.
Observations of CO in dwarf galaxies have been very challenging and
the glaring underabundance of observed CO in dwarf galaxies and
relatively high FIR/CO luminosities have often been interpreted as
unusually high star formation efficiency.  While all these
observational effects are a consequence of the lower metal abundance
and decreased dust to gas ratio, we do not find an unambiguous direct
correlation of the I[CII]/I(CO) and I[CII]/FIR ratios in our surveys
with metallicity.
The reduced dust abundance in these environments allows the UV
radiation to penetrate deeper, leaving a smaller CO core surrounded by
a larger C$^{+}$- emitting region, thus enhancing the I[CII]/I(CO)
ratios \cite{maloney88}. Consequently, as the FUV flux travels further,
the intensity becomes geometrically diluted, resulting in a lower
beam-averaged FIR flux, accounting for the increased I[CII]/FIR ratios
\cite{israel96}.

Using the results of recent PDR models that consider the effects of
reduced metallicity \cite{kaufman99}, we can find solutions for the
dwarf galaxies for clouds in our beam described by 2 different
cases. One possible solution (case A) is for clouds with low A$_{v}$
($\sim$3) and equal densities (n) in the CO and C$^{+}$- emitting
regions with n ranging from 10$^{3}$ to 10$^{4.5}$ cm$^{-3}$ and low
to moderate G$_{o}$ (normalized to the local interstellar radiation
field intensity, 1.3x10$^{-4}$ erg s$^{-1}$ cm$^{-2}$ sr$^{-1}$)
ranging from 10$^{1.5}$ to 10$^{3}$. Another possible solution (case
B) is a higher A$_{v}$ ($\sim$10) with the density of the CO-emitting
region (n$_{CO}$) $>$ the density of the C$^{+}$- emitting region
(n$_{CII}$) which gives higher ranges of G$_{o}$ ($\sim$ 10$^{2.5}$ to
10$^{3.5}$). We can put further constraints on these solutions through
stellar population modeling. Based on our modeled SED for IIZw40, for
example, case A is a solution (section 4.1). Arguments for molecular cloud stability
point toward case B for the LMC \cite{kaufman99}. Decreasing the A$_{v}$
(case A) or increasing the n$_{CO}$ relative to n$_{CII}$ (case B) has
the similar effect of reducing the CO-emitting core and increasing the
C$^{+}$- emitting zone and increasing the CO-to-H$_{2}$ conversion
factor
\cite{kaufman99}.  Based on [CII] observations in IC10, for example, we speculated
that up to 100 times more H$_2$ may be `hidden' in a C$^+$-emitting
regions compared to that deduced only from CO observations and using
the Galactic CO-to-H$_{2}$ conversion factor \cite{madden97}.  The
presence of H$_2$ in the C$^+$- emitting region is due to the
self-shielding of H$_2$ from UV photons or shielding by dust
\cite{burton90}
\cite{pak98} \cite{kaufman99}. 
\section{Mid-Infrared Observations}
\begin{figure*}
\includegraphics[scale=.6,viewport=0 0 630 630, clip]{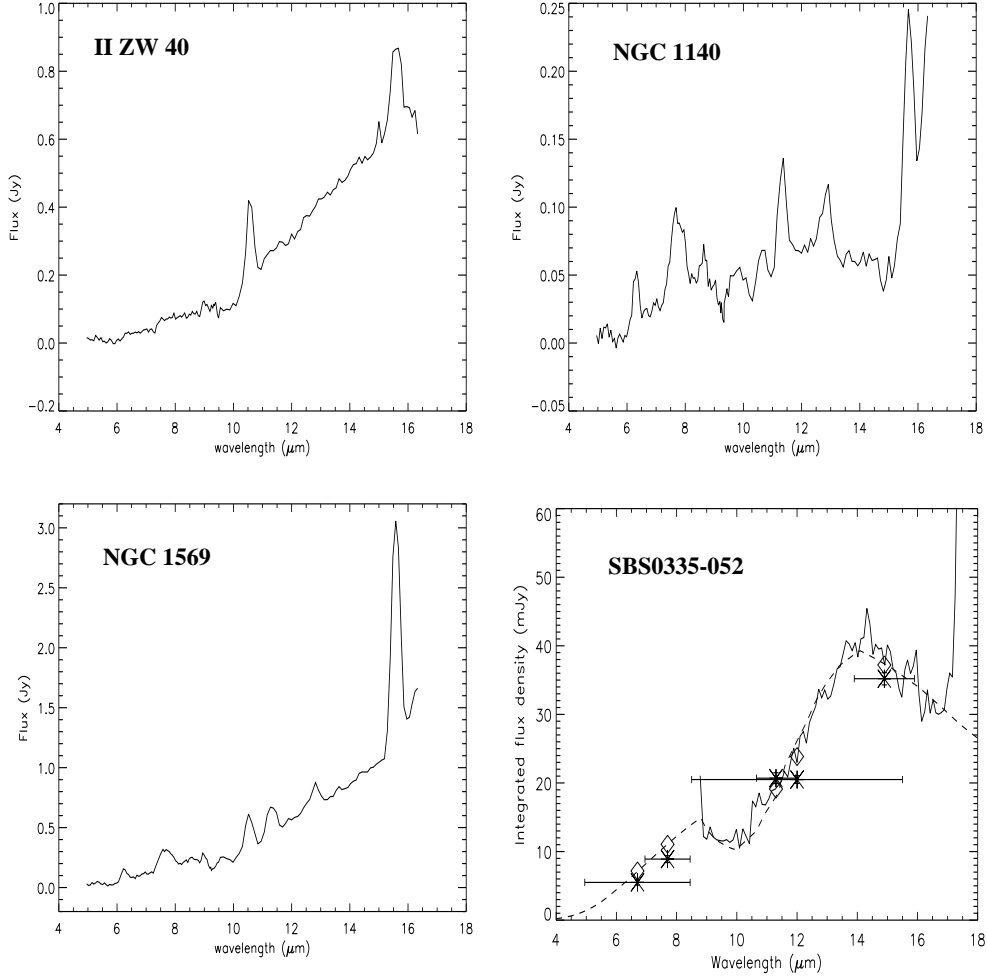}
\caption{ISOCAM MIR spectra of dwarf galaxies: IIZw40, NGC~1569, NGC~1140 and SBS0335-052. The horizontal lines in SBS0335-052 are broad band measurements; the dashed line is a blackbody with  A$_{v}$ $\sim$ 20 \cite{thuan99a}. Note the absorption attributed to amorphous silicates at $\sim$ 9 and 18 $\mu$m.}
\label{cvfs}
\end{figure*}

We are studying some of these galaxies in our [CII] survey with
followup MIR observations.  In Figure \ref{cvfs} we show ISOCAM
\cite{ccecarsky96} spectra covering 5 to 17 $\mu$m for 3 galaxies from
our [CII] survey, IIZw40, NGC~1140 and NGC~1569 along with that of
the notoriously metal-poor SBS0335-052 \cite{thuan99a}. The
spectra represent the total emission from the galaxies except in the
case of the NGC~1569 spectra, which samples the region around the
H$\alpha$ peak $\#$2 (see
\cite{waller91}).  As often seen in starburst galaxies
(e.g. \cite{dudley99} \cite{laurent00}), the MIR spectra are dominated
by steeply rising continua longward of $\sim$ 10 $\mu$m, as evident in
NGC~1569, IIZw40 and SBS0335-052 (Figure \ref{cvfs}).  Thermal
emission from hot small grains with mean temperatures of the order of
100's of K are responsible for the MIR continuum emission. The
unidentified infrared bands (UIBs) at 6.2, 7.7, 8.6, 11.3 and 12.6
$\mu$m, are proposed to be due to aromatic hydrocarbon particles
undergoing stochastic temperature fluctuations (i.e, PAHs
\cite{leger84} \cite{allaman89}; coal grains
\cite{papou91}) and are observed to peak on the PDR zones around
the HII regions but are destroyed deep within HII regions
\cite{vertraete96} \cite{dcesarsky96} \cite{tran98}. While the UIBs are not obvious in the spectra of IIZw40 and SBS0335-052, and are only very weakly present NGC~1569, they can be distinguished in the spectrum of NGC~1140.  Several ground state fine-structure
nebular lines are present also in 3 of the spectra, the most prominent
being 15.6 $\mu$m [NeIII] (energy potential $\sim$ 41 eV) and 10.5 $\mu$m [SIV] (energy potential $\sim$ 35 eV).
Weaker, lower energy lines may also present, such as the 8.9 $\mu$m
[ArIII] line and the [NeII] 12.8 $\mu$m line, which can be blended
with the 12.6 $\mu$m UIB.  All of these spectra look very
different from one another and all differ significantly from those of
normal metallicity starburst galaxies. Normal starburst galaxies show
prominent UIBs, in contrast to AGNs, which are devoid of UIBs
(e.g. \cite{roche91} \cite{dudley99} \cite{laurent00}). When compared to spectra characteristic of PDRs and HII regions, ie, M17 \cite{dcesarsky96} \cite{vertraete96}, IIZw40 is remarkably similar to that of an HII region. In contrast,
NGC~1140, which has a very flat continuum, yet very strong [NeIII]
line, does have a more obvious contribution from PDR regions in its
spectra. The MIR spectra of N66, the most prominent HII region in the
SMC, also shows a scarcity of UIBs in the vicinity of the most massive
central cluster \cite{contoursi99}, as does the low metallicity source
NGC~5253 \cite{crowther99}. 
In some starburst galaxies, amorphous silicate is seen in absorption
centered at 9 and 18 $\mu$m (e.g. \cite{roche91}, \cite{dudley99},
\cite{laurent00}).  We can fit the MIR region of the IIZw40 
spectrum with a blackbody of 193 K and and an absorption equivalent to
A$_{v}$ $\sim$ 4. We caution interpretation of the dust temperature we
derive assuming a blackbody, since the dust emitting in the MIR is
expected to be undergoing stochastic heating events, rather than being
in thermal equilibrium with the radiation field. The amount of
absorption in IIZw40 (A$_{v}$ $\sim$ 4) has yet to be confirmed. In
SBS0335-052, a very low metallicity galaxy (1/40 solar), A$_{v}$ $\sim$
20 deduced from the absorption in the ISOCAM MIR spectra (Figure \ref{cvfs})
\cite{thuan99a}.  The presence of a significant amount of dust in
such a low metallicity galaxy is surprising, since star formation in SBS0335-052 began as recently as 100 Myr ago \cite{papader98}
\cite{thuan99b}. Such high extinction implies that the current star
formation rate, hidden by dust, can be underestimated by at least 50\%
\cite{thuan99a}
\subsection{Effects of the starburst activity on the dwarf galaxy MIR spectra}
As a consequence of the decreased dust abundance in dwarf galaxies, the
ISM throughout the galaxies is effected globally by the hard radiation
field of the massive stellar clusters. These galaxies contain evidence
for Wolf-Rayet stars {\cite{schaerer99} and super star clusters have
been detected in NGC~1140 \cite{hunter94}, NGC~1569 \cite{oconnell94}
and SBS0335-052
\cite{thuan97}. The harsh radiation field, which more easily permeates the
ISM compared to normal metallicity environments, can destroy the UIB
carriers, for example, over very extensive spatial areas.  The effect of the
pervasive radiation field can be witnessed in NGC~1569 (Figure
\ref{n1569_image}).  Photodissociation occurs on global scales. Violent activity is revealed by the H$\alpha$ distribution \cite{waller91} \cite{martin98} and the 15.8 $\mu$m [NeIII] emission,
with giant streamers suspected to originate from the energetic winds
of the super star clusters A \& B, (shown in the figure as white
stars). The UIBs, [SIV] and [NeIII] emission seem to avoid the super
star clusters, which blow out much of the gas and dust on relatively
short time scales. This effect is also seen in the CO \cite{taylor99},
HI \cite{israel90} and the H$\alpha$ \cite{waller91}
distribution. Likewise we see the destruction of the UIBs in the beam-
averaged spectrum of the entire galaxy of IIZw40 and SBS0335-052 (due
to our lack of spatial resolution we do not see the details within
these galaxies in the MIR).
\begin{figure*}
\begin{center}
\includegraphics[scale=0.6,viewport=97 176 471 571, clip]{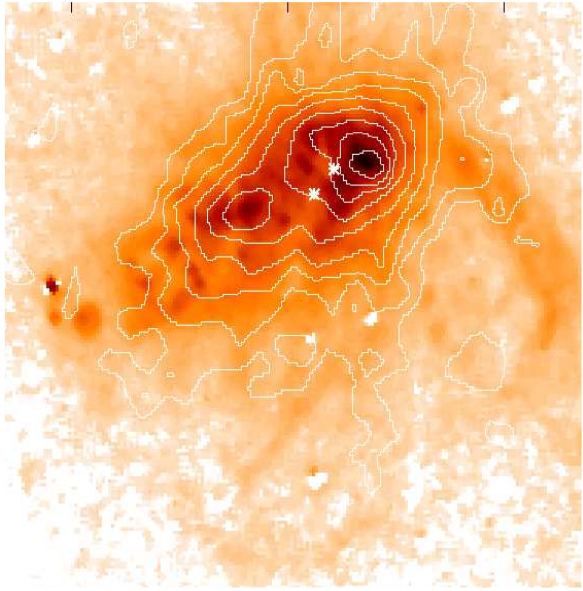}
\caption{NGC~1569: H$\alpha$ (image) \cite{waller91} and 15.8 $\mu$m [NeIII] emission (contours). Note the extended [NeIII] filaments as also seen in H$\alpha$. The 2 white stars mark the positions of the super star clusters A and B \cite{oconnell94} which are void of[NeIII] emission.}
\label{n1569_image}
\end{center}
\end{figure*}
\section{Spectral Energy Distribution}
We compile broad-band data from the literature for IIZw40, NGC~1569
and NGC~1140, and together with our MIR data, construct  stellar
spectral energy distributions (SEDs). In doing so, we fit the observed optical
and NIR data with stellar evolution models of PEGASE \cite{fioc97},
taking into account the results of photoionization
modeling of the MIR line emission using CLOUDY \cite{ferland96}. This
is an attempt to reconstruct the input stellar spectra consistent from
the viewpoints of both the stellar evolution and
photoionization.
\subsection{Combined stellar evolution and photoionization model results} 
Using PEGASE with an instantaneous star formation rate, metallicity
0.2 solar and a Salpeter IMF (with upper and lower mass cut offs of .1
and 120 solar masses), we find solutions to observed broad band
stellar light for various ages and ionization parameters. Diagnostic
optical and NIR lines in the literature exist for all of these sources for a
variety of apertures. The ISOCAM MIR observations also provide
important diagnostic lines of neon, sulphur and argon, and has been
recently addressed by others, including \cite{lutz98}
\cite{crowther99} \cite{schaerer99b}
\cite{genzel98}.  For example, the [NeIII]/[NeII] ratio, is a measure of T$_{eff}$, the hardness of the radiation field, and therefore traces the massive
stellar population.  For the dwarf galaxies, we find [NeIII]/[NeII]
ratios in the range of 5 to 10 - much higher than those for normal
metallicity galaxies ($\leq$1) \cite{thornley00}. The extreme values
of the [NeIII]/[NeII] ratios are due to effects of the
low-metallicities of the systems: the T$_{eff}$ of the stars increases
as the metallicity decreases for a specific stellar age.  High ratios
of [NeIII]/[NeII] and the prominent [SIV] in these spectra limit the
age of the present star formation to $<$ 5 Myr.  Beyond this age, the
massive stars have died and the [NeIII]/[NeII] ratio drops
dramatically. The high excitation 24.9 $\mu$m [OIV] line, covered by
the ISO SWS data, is observed in some dwarf galaxies
\cite{lutz98} and has been proposed to be due to the presence of Wolf-Rayet
stars \cite{schaerer99b}. For NGC~1569, NGC~1140 and IIZw40, we
construct composite stellar SEDs that require a 75\% to 95 \% mass
fraction of an 'older' population ranging in age from about 10 Myr to
30 Myr along with 5\% to 30\% of a very young population, $<$ 5
Myr. The broad band optical and NIR data alone reveal predominantly
the older population in our apertures. Figure \ref{sed}
shows an example of the resultant composite SED obtained for IIZw~40,
and the extreme-ultraviolet (EUV) radiation which the young, massive
stellar population traces. Observational evidence for the presence of
Wolf-Rayet stars also indicates a very young stellar population \cite{vacca92}.
\begin{figure*}
\includegraphics[width=\textwidth, viewport=0 0 670 495, clip]{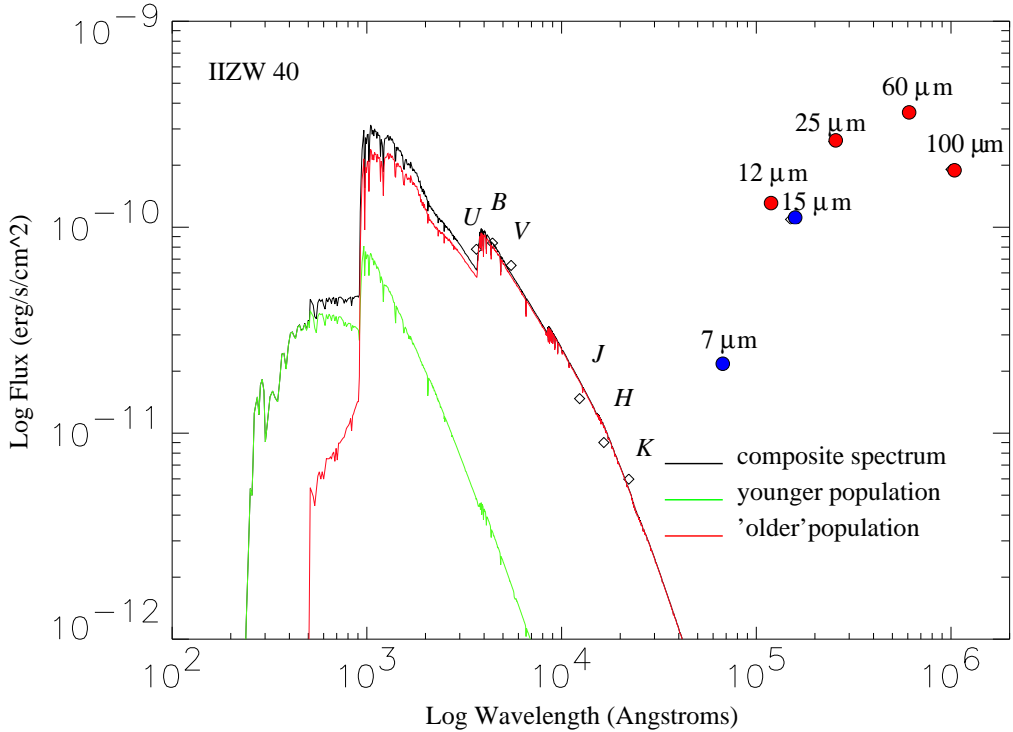}
\caption{IIZw40 SED. The synthetic stellar spectra are fit for the extinction-corrected optical to NIR data from the literature for a 12'' aperture using PEGASE. The 12, 25, 60 and 100 $\mu$m data are from IRAS and the 7 and 15 $\mu$m data points are integrated over 5.0 to 8.5 $\mu$m and 12.0 to 17 $\mu$m bands, respectively, using the ISOCAM spectrum (Figure \ref{cvfs}).}
\label{sed}
\end{figure*}
\section{Dust modeling}
Having modeled the radiation field above, we next use the stellar
spectra of IIZw40, NGC~1569 and NGC~1140 as input to a dust model to
deduce the nature of the various dust components emitting in the MIR
and the FIR. This is an important step since dust plays a major role
in influencing the chemical and physical state of the ISM. We use the
D\'esert et al. model
\cite{desert90}, which calculates the IR emission from large silicate
grains (BGs), very small amorphous carbon grains (VSGs), and
stochastically-heated polycyclic aromatic hydrocarbons (PAHs), for
various grain size distributions. This model is rather empirical in
its approach and thus does not give an exact fit to the details of the
observed spectrum. For example, the 8.6~$\mu$m UIB is not well-matched
and no emission from bands at wavelengths longer than 11.3~$\mu$m are
included. The model is currently in the process of modification using
up-to-date laboratory-measured optical constants for a wide range of
likely interstellar grain materials.
\subsection{Dust in low-metallicity galaxies}
In Figure \ref{dust_model} we show, as an example, the ISOCAM MIR
spectrum and the IRAS data points for IIZw40 and NGC~1569, where we
have plotted the emission from the PAH (dashed line), VSG (dotted
line) and BG (dashed-dotted line) components. In these galaxies the
MIR spectrum is clearly dominated by emission from VSGs with very
little PAH emission. The BG component dominantes the overall dust
emission with mass fractions ranging from 93\% to 99\% for the 3
galaxies, while the PAH mass fraction is relatively insignificant - 5
orders of magnitude lower. This model gives a PAH/VSG mass ratio for
NGC~1569 and IIZw 40 of 2 to 3x10$^{-4}$ and 10 times this for
NGC~1140. The D\'esert et al. model applied to the Galactic cirrus
gives PAH/VSG mass ratio $\sim 1$. Thus, even compared to the VSG
population, we find an insignificant mass fraction of PAHs, reflecting
the fact that the PAHs are destroyed in the hard radiation fields
in these galaxies. PAHs are thought to be the primary particles
responsible for the photoelectric heating process \cite{bakes94} and
are incorporated in PDR models
\cite{kaufman99}. Our preliminary results, while not statistically
robust at this stage, suggest that even in the absence of PAHs, the
photoelectric effect is efficient, as both IIZw40 and NGC~1569 are
relatively prominent [CII] sources from our survey. On the contrary,
in NGC~1140, where PAHs are more obvious in the MIR spectra (Figure
\ref{cvfs}), we do not detect [CII]. VSGs (sizes determined from model $\sim$40 to 300A), which are very abundant relative to the PAHs in NGC~1569 and IIZw40, and less so in NGC~1140, may therefore, be the more efficient
sources of photoelectric gas heating in these environments, rather
than PAHs. More detailed studies of these galaxies will be carried out
using the analytical dust model of V\'arosi and Dwek \cite{varosi99},
which takes into account radiative transfer in a two-phase clumpy
environment and considers various geometries.
\begin{figure*}
\begin{center}
\includegraphics[width=\textwidth,viewport = 23 307 716 590,clip]{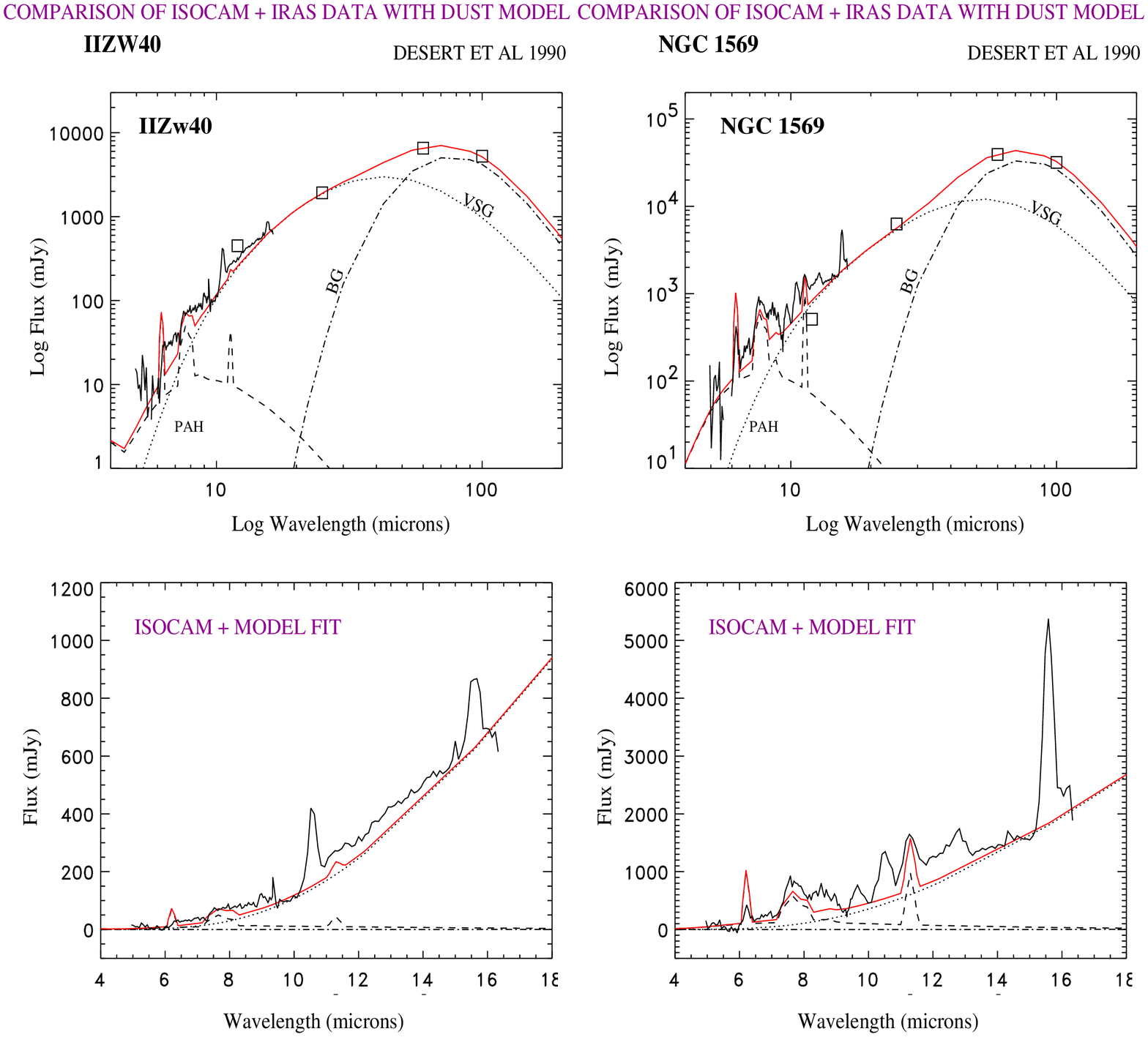}
\caption{Model results for the dust components in IIZw40 (left) and NGC~1569 (right) fitted to the ISOCAM MIR spectra and the IRAS data (boxes) Component are from the D\'esert et al. model \cite{desert90} (see text for model explanation)}
\end{center}
\label{dust_model}
\end{figure*}
\section{Summary}
Tracers of various components of the ISM show evidence of effects of
the hard stellar radiation field in dwarf galaxies on the
surrounding ISM due to the decreased dust abundace, allowing
photoionization over large galactic scales to occur. From our survey
of the 158 $\mu$m [CII] PDR cooling line in dwarf galaxies, we observe
an increased penetration of the FUV radiation field which enhances the
I[CII]/I(CO) emission in dwarf galaxies up to a factor of 10 more than
in normal metallicity star burst galaxies. We also find a small
enhancement in the I[CII]/FIR ratio in dwarf galaxies. Followup MIR
ISOCAM spectroscopy provides details of ionic lines, UIBs and small
hot small grain emission distribution in dwarf galaxies. The strong
MIR [NeIII]/[NeII] ratios are signatures of the hard radiation fields
and indicate the presence of young massive stellar populations in
dwarf galaxies. Because of the increase in T$_{eff}$ in low
metallicity environments, this ratio is enhanced at least 5 to 10
times more in dwarf galaxies than in normal metallicity galaxies. The
penetrating radiation field also effects the dust components,
destroying the UIBs in some dwarf galaxies on global scales, as
is evident in the MIR spectra and in the dust modeling.
\section{Acknowledgements}
This work, still in progress, results from a series of observations
from ISO and the KAO and includes a number of collaborators such as
S. Colgan, N. Geis, M. Haas, D. Hollenbach, A. Jones, P. Maloney,
A. Poglitsch, D. Ragaigne, B. Smith and M. Wolfire. I have benefited
from invaluable discussions with E. Dwek and A. Jones on dust
modeling. I thank W. Waller for his H$\alpha$ image of NGC 1569.

\end{document}